\documentclass[twocolumn,aps,prl,showpacs,superscriptaddress]{revtex4}
\usepackage{graphicx}
\usepackage{amssymb} 
\usepackage{url}     
\usepackage{bm}      
\usepackage{color}

\newcommand{\bise}{Bi$_2$Se$_3$}

\newcommand{\alum}{Al$_2$O$_3$}
\newcommand{\haf}{HfO$_2$}
\newcommand{\vtg}{$V_{TG}$}
\newcommand{\esh}{$e^2/h$}
\newcommand{\RH}{$R_{H}$}
\newcommand{\lphi}{$l_{\phi}$}
\newcommand{\tauphi}{$\tau_{\phi}$}

\newcommand{\delG}{$\Delta G$}
\newcommand{\mc}{magnetoconductance}

\begin{document}

\author{H. Steinberg}
\affiliation{Department of Physics, Massachusetts Institute of Technology, Cambridge, MA 02139, USA}
\author{J. -B. Lalo\"{e}}
\affiliation{Francis Bitter Magnet Lab, Massachusetts Institute of Technology, Cambridge, Massachusetts 02139, USA}
\author{V. Fatemi}
\affiliation{Department of Physics, Massachusetts Institute of Technology, Cambridge, MA 02139, USA}
\author{J. S. Moodera}
\affiliation{Department of Physics, Massachusetts Institute of Technology, Cambridge, MA 02139, USA}
\affiliation{Francis Bitter Magnet Lab, Massachusetts Institute of Technology, Cambridge, Massachusetts 02139, USA}
\author{P. Jarillo-Herrero}
\affiliation{Department of Physics, Massachusetts Institute of Technology, Cambridge, MA 02139, USA}


\title{Electrically tunable surface-to-bulk coherent coupling in topological insulator thin films}

\begin{abstract}
We study coherent electronic transport in charge density tunable micro-devices patterned from thin films of the topological insulator (TI) \bise. The devices exhibit pronounced electric field effect, including ambipolar modulation of the resistance with an on/off ratio of 500\%. We show that the weak antilocalization correction to conductance is sensitive to the number of coherently coupled channels, which in a TI includes the top and bottom surfaces and the bulk carriers. These are separated into coherently independent channels by the application of gate voltage and at elevated temperatures. Our results are consistent with a model where channel separation is determined by a competition between the phase coherence time and the surface-to-bulk scattering time.
\end{abstract}

\maketitle

\date{\today}
Topological Insulators (TIs) are gapped bulk insulators with gapless Dirac surface states which have emerged as a new paradigm in the study of topological phases of matter~\cite{TI_General}. TI-based electronic devices are attractive as platforms for spintronic applications~\cite{Yazyev_2010}, and for detection of emergent properties such as Majorana excitations~\cite{Fu_Kane_Majorana_2008}, electron-hole condensates~\cite{Seradjeh_2009} and the topological magneto-electric effect~\cite{Qi_TME_2009}. Most theoretical proposals envision an experimental geometry consisting of a planar TI device, where electrical current is carried by the surface states. 

Despite considerable recent evidence of TI surface states in ARPES~\cite{Hsieh_2008,TI_ARPES} and STM~\cite{STM_cite}, their observation and manipulation in transport experiments remains difficult: TI devices require a density-tunable surface state which is decoupled from the residual bulk carriers. Realistic devices, however, conduct through parallel channels consisting of the top and bottom surfaces, and of bulk carriers which, due to unintentional doping, can account for a significant part of the conductance, limit the surface density tunability, and create an uncertainty in the surface-to-bulk coupling. It is therefore desirable to minimize the bulk contribution and to simultaneously investigate the various ways in which it is involved in electronic transport. Suppression of the bulk channel can be obtained by minimizing carrier density, as demonstrated in mm-size single-crystals~\cite{Analytis_Qu}, and by fabrication of nanoscale devices, such as nanoribbons~\cite{Peng_2010} and flakes exfoliated from single-crystals~\cite{Steinberg_2010,Sacepe_2011,Checkelsky_2011}. The latter studies have also demonstrated that the surface carrier density can be tuned by electrostatic gating and have all detected an ambipolar modulation of conductance. 

\begin{figure}
\label{Figure_1}
\begin{center}
\includegraphics[width = 86mm]{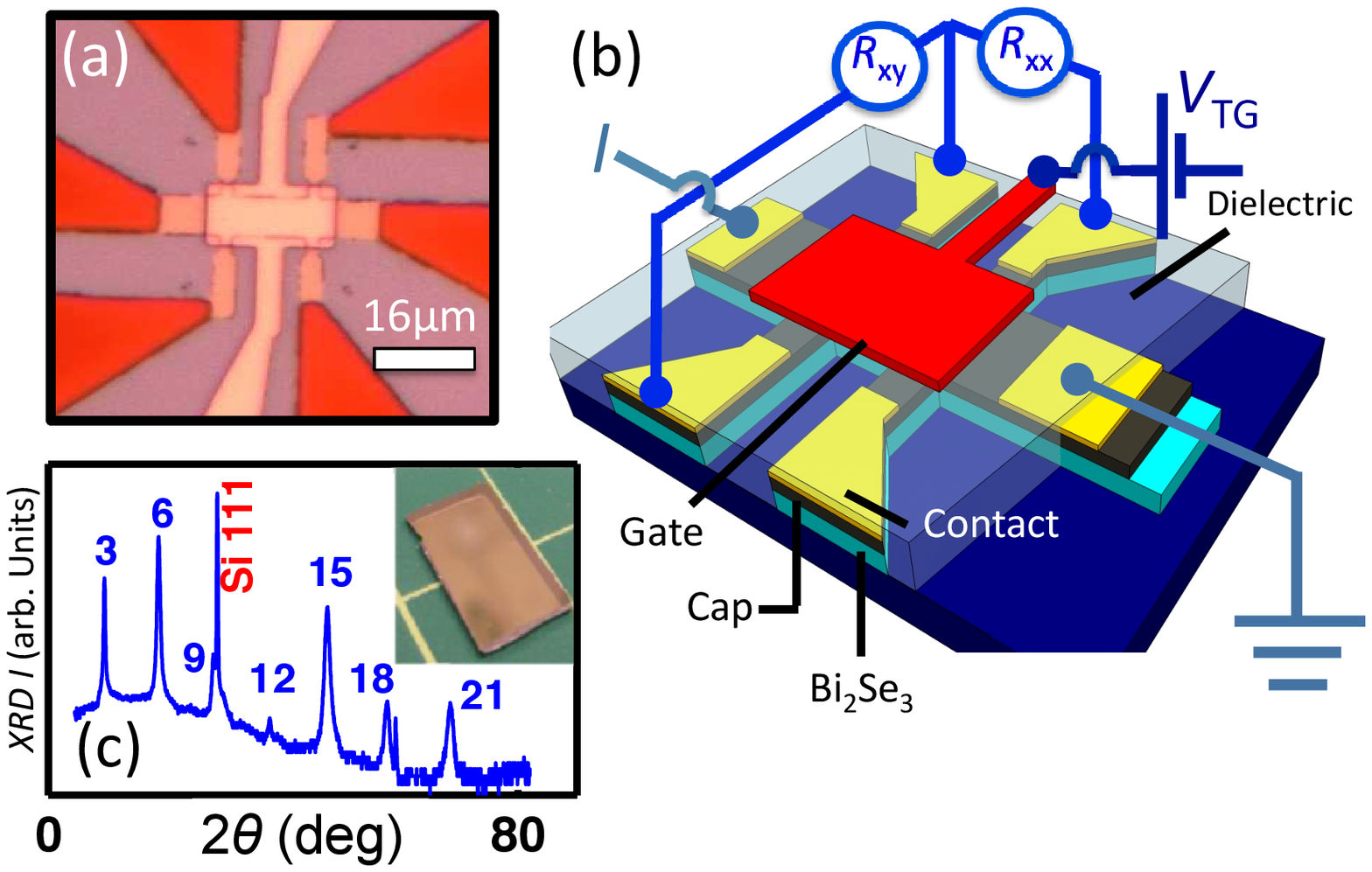}
\vspace{-0.8cm}
\caption{(a) Optical image of a gated \bise\ device, showing a \bise\ Hall bar, evaporated contact pads, and top-gate. The gate electrode covers the entire area of the Hall bar. (b) Schematic of the device and measurement circuitry, showing the \bise\ layer, cap layer, contact pads, dielectric, and top gate. (c) X-Ray Diffraction intensity of a \bise\ thin film, showing the (0,0,3n) family of diffraction lines indicating growth along the \textit{c} axis. Some lines associated with the Si substrate are also visible. Inset: Image of \bise\ film grown on a Si wafer.}
\end{center}
\vspace{-0.7cm}
\end{figure}

An alternative route for fabrication of thin TI devices employs the growth of thin films in ultra-high vacuum chambers. Thin film growth offers fine control over geometry and composition and a straightforward approach for growth of heterostructures. High quality \bise\ thin films were studied by in-situ STM~\cite{Cheng_LL_MBE_2010} and ARPES~\cite{YiZhang_2010}. Transport studies~\cite{MBE_WAL,Chen_GateWAL_2010} carried out on the same films are still dominated by bulk transport since the material is highly doped. Interestingly, most thin film studies report a pronounced weak antilocalization (WAL) feature, which is an indication of phase coherent transport.

WAL and weak localization emerge from the correction of coherent time-reversed closed paths to electronic transport~\cite{Bergmann_1984}. They are sensitive to the competition between the phase coherence time \tauphi\ and other time-scales, and have been extensively employed as a probe for coherent dynamics in solid state systems~\cite{Beenakker_vanHouten_1991}. In TIs both the bulk and surface states may contribute to coherent transport, and in both cases they should exhibit WAL: In the bulk, the strong spin-orbit (SO) coupling leads to random rotations of the spin orientation~\cite{Bergmann_1982}, which on average result in destructive interference for backscattering paths, hence leading to enhanced conductance. On the chiral TI surface state, momentum is coupled to the spin degree of freedom, so time-reversed paths around a closed trajectory acquire a relative phase of $\pi$~\cite{McCann_2006}.  

Given the various parallel conduction channels, so far it was not clear why most of the recent TI transport studies report WAL corrections corresponding to a single coherent channel~\cite{MBE_WAL,Chen_GateWAL_2010}, and it was suggested that one of the surfaces has a significantly higher coherence length than the other~\cite{Chen_GateWAL_2010}. A few studies have found the number of channels to be tunable by a gate voltage~\cite{Checkelsky_2011, Chen_GateWAL_2010, Chen_2011_WAL}, but the microscopic mechanism underlying this tunability was not thoroughly investigated. Here we study the WAL effect in charge density tunable \bise\ devices. We find that the number of independent coherent channels is tunable by the applied electric field and by the temperature. Our results indicate that bulk carriers play a crucial role in TI coherent transport, and that channel coupling is controlled by a competition between the phase coherence time and the surface-to-bulk coupling time.


\begin{figure}
\label{Figure_2}
\begin{center}
\includegraphics[width = 74mm]{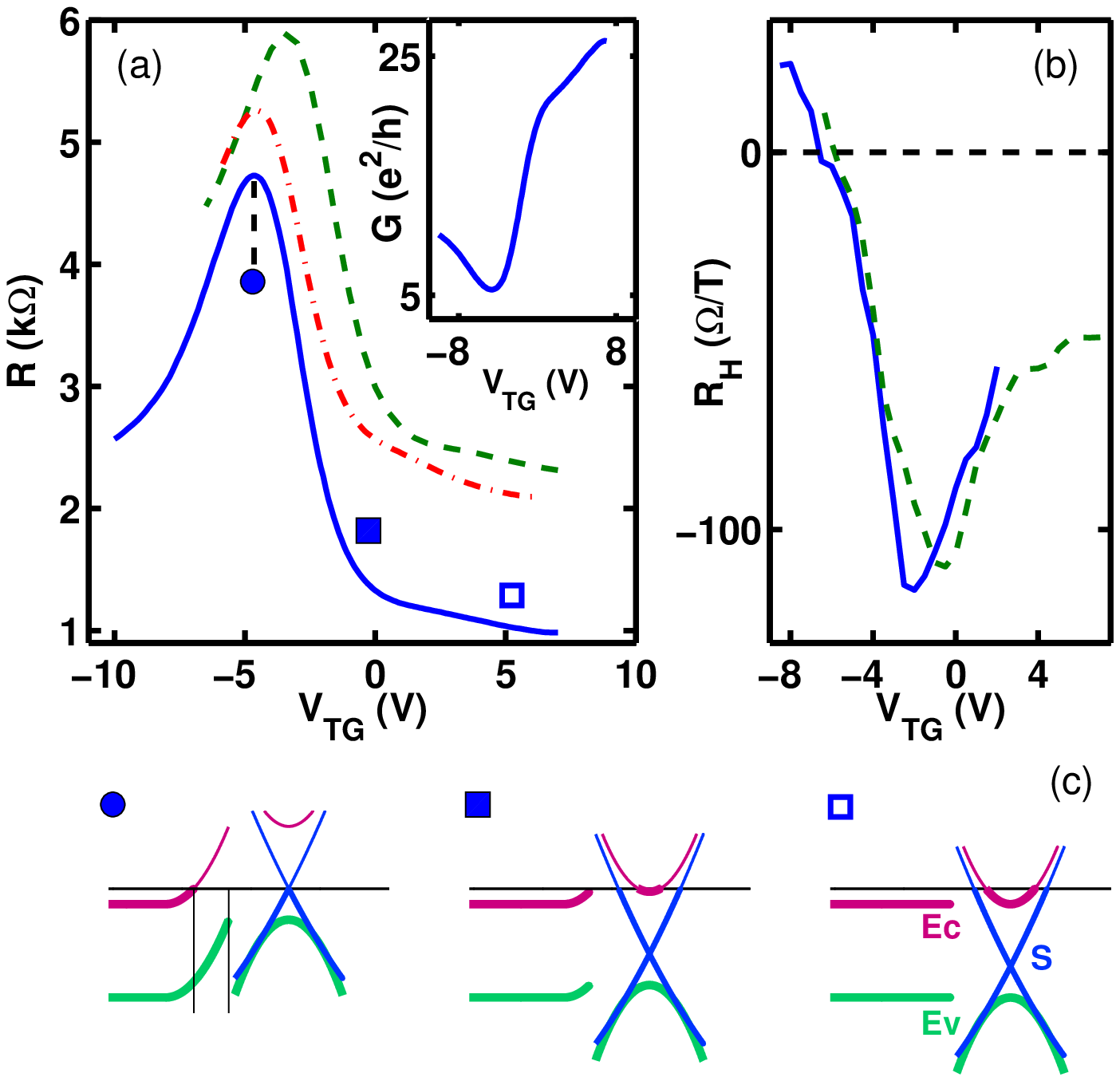}
\vspace{-0.6cm}
\caption{(a) Square resistance $R$ vs.~top-gate voltage \vtg\, measured on devices A1 (continuous line), A2 (dashed) and A3 (dash-dot) at $T$ = 4K. Inset: Square conductance, $G$, for device A1. (b) Low field Hall coefficient \RH\ in 2D units. \RH\ is observed to cross over to positive values at \vtg\ $\sim$ -7V. (c) Schematics of the band structure and spatial variation of bulk bands near the surface at different gate voltages. In each subpanel, the left side shows the band energy vs. vertical position, and the right side is the energy-momentum dispersion at the surface. The bulk conduction band is marked by "Ec"; Valence band by "Ev"; and surface by "S". Right  panel (hollow square): \vtg\ $>$ 0. The bulk states near the surface are populated. Middle panel (solid square): Bulk states near the surface are at the depletion point (bottom of conduction band). Left panel (circle): \vtg\ $<$ 0. Top surface state is gated to the Dirac point. The bulk bands bend to form a depletion layer near the surface, marked by vertical lines. This depletion region increases with increasing negative $V_{TG}$. The markers in panel (a) correspond to the band structure schemes shown in the bottom panels.}
\end{center}
\vspace{-0.9cm}
\end{figure}

We fabricate low density \bise\ devices by growing large area thin films on a Si(111) substrate and subsequent lithographic processing~\cite{SI}. A device image and schematics are shown in Figs.~1(a) and (b). X-Ray diffraction (Fig.~1(c)) reveals sharp (0,0,3n) peaks, confirming that the films are \textit{c}-axis oriented along the growth direction. The \bise\ is typically capped in-situ by a sputtered layer of 4nm \alum. Additional 16nm of \haf\ gate dielectric are deposited after device patterning.

Figure~2 shows the electric field effect behavior of three representative devices (Devices A1,2,3), all patterned from the same 20nm thick capped \bise\ wafer, and measured at $T$ = 4K. Every resistance (conductance) shown is resistance (conductance)-per-square. For \vtg\ $>$ 0, $R$ is modulated weakly by the gate, but begins rising sharply around \vtg\ = 0, and peaks at 5 - 6 k$\Omega$ around \vtg\ = -5V. The inset shows the square conductance $G$ of Device A1 in units of \esh. For this device $G$ varies between 5 and 25 \esh\ , exhibiting a pronounced ambipolar modulation characteristic of Dirac dispersions, and indicating that the conductance is dominated by the top surface. This is supported by the Hall coefficient \RH, which for a single channel is given by \RH~$= -1/ne$, where $n$ is the density of charge carriers and $e$ is the electron charge, but which is more complex for multi-channel systems~\cite{Steinberg_2010}. \RH\ is strongly modulated by the gate and crosses over to positive values (Fig.~2(b)) at \vtg\ $\sim$ -7V, indicating that the dominant Hall current-carrying population has changed from electrons to holes, as expected for Dirac systems with low doping. This is in stark contrast to previously reported studies on \bise\ thin films, which are very strongly electron-doped, and do not exhibit such a strong carrier type modulation~\cite{MBE_WAL,Chen_GateWAL_2010}. Gating of the surface states shifts the surface energy bands vertically and hence is accompanied by bending of the bulk bands near the surface, as shown in Fig.~2(c). We associate the sharp change in the slope of $R(V_{TG})$  near \vtg\ = 0 with the depletion of bulk carriers immediately near the surface. For \vtg\ $<$ 0, the main effect of the gate is to change the TI surface state charge density, which results in a more rapid variation of conductance. However, any change of the surface charge density has to be accompanied by a change in the width of the bulk depletion region. This depletion region is crucial for coherently separating the surface state from the bulk bands.

We now turn to the electric field effect on coherent transport, studied by \mc. These data were taken on Device B, which is patterned from a different wafer than the one reported above but has similar transport characteristics, with a well defined resistance maximum (Fig.~3(b)). Figure~3(a) shows $R$ vs.~perpendicular magnetic field, $B$, at $T$ = 0.3K and where WAL appears as a sharp suppression of resistance at low magnetic field. We repeat the same measurement for different gate voltages, and find that the WAL feature also evolves with \vtg, as shown in Fig.~3(c), where the change in conductance, $\Delta\textit{G}(B) = G(B) - G(0)$, is plotted at \vtg\ = +6V and \vtg\ = -8V. Both traces exhibit a correction of $\Delta G \sim e^2/h$, although the former is sharper in magnetic field. The WAL correction depends on the phase-coherence time \tauphi, or length \lphi, which are related through $l_\phi = \sqrt{D\tau_\phi}$ ($D$ being the diffusion constant). The data agree well with~\cite{HLN_1980}: 
\begin{equation}\label{HLN}
\Delta G_{WAL}(B) = -\alpha \frac{e^2}{2\pi^2\hbar}\left[\log{\frac{B_o}{B}}-\psi\left(\ \frac{1}{2}+\frac{B_o}{B}\right)\ \right]
\end{equation} where \textit{B}$_o$ = $\hbar/(4el_\phi^2)$, $\psi$ is the digamma function, and $\alpha$ is a prefactor which should be equal to -1/2 for a single coherent channel. Eq.~\ref{HLN} is valid for massless Dirac fermions (i.e. the surface states)~\cite{McCann_2006} and is also valid in the bulk where the spin-orbit scattering time $\tau_{SO}$ is significantly shorter than \tauphi~\cite{SI}. The data (blue) agree well with the fit to Eq.~\ref{HLN} (yellow) over the entire measured range in both cases, indicating that WAL constitutes the entire correction to \mc. The fit contains two free parameters, the coherence length \lphi\ and the prefactor $\alpha$, both of which change with the applied gate voltage (Fig.~3(d)). Here we focus on the gate-tunability of $\alpha$, which changes from -0.7 to $\sim$ -1, reflecting a change in the effective number of coherent channels. 


\begin{figure}
\label{Figure_3}
\vspace{-0.5cm}
\begin{center}
\includegraphics[width = 78mm]{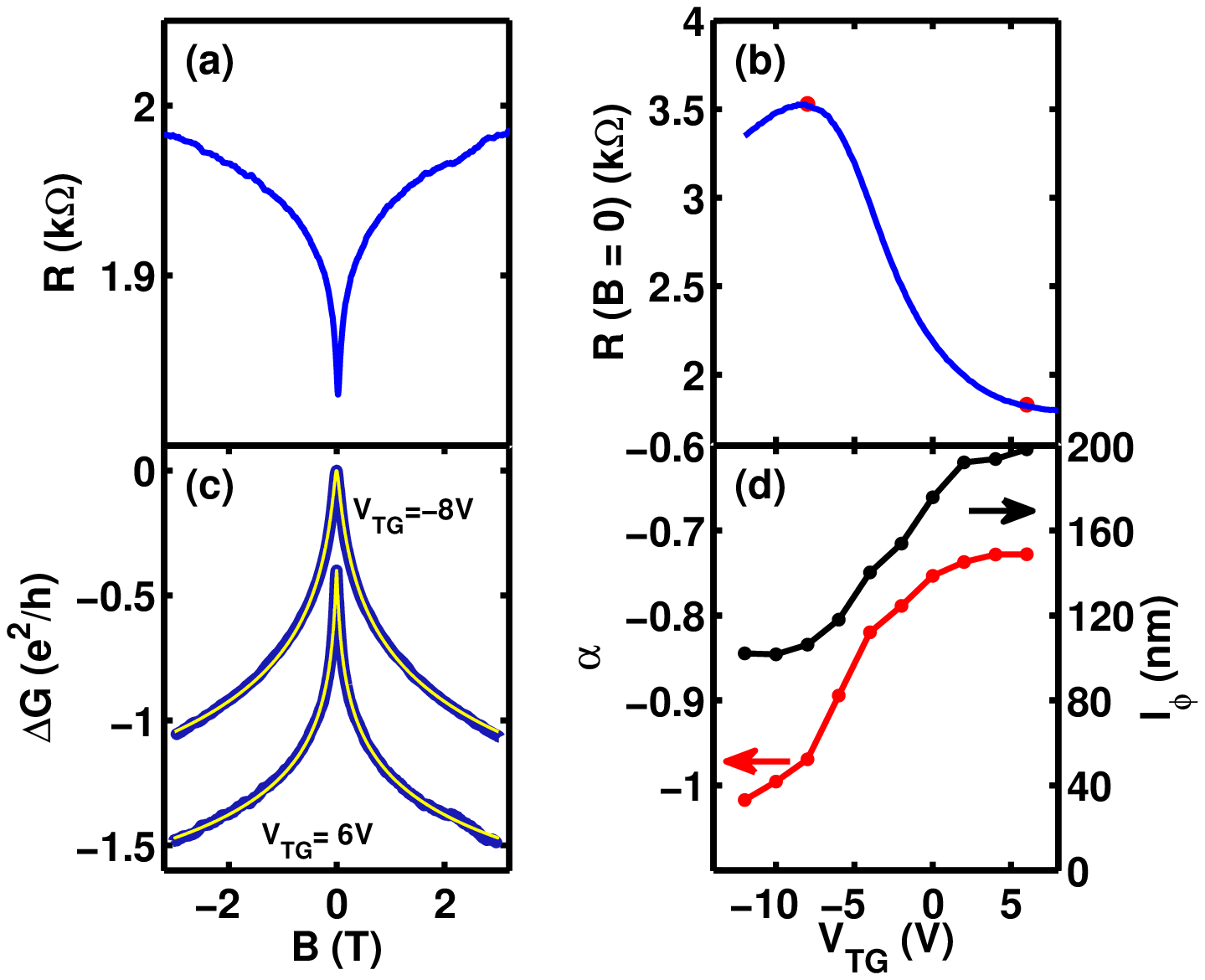}
\caption{(a) \mc\ of Device B taken at \vtg\ = +6V showing a pronounced WAL suppression of resistance. (b) \textit{R}(\textit{B} = 0) vs. \vtg. \textit{R} peaks at \vtg\ = -8V. (c) \delG\ vs. \textit{B} plotted for \vtg~= +6V and -8V (marked by circles in (b)). The two plots are offset for clarity. The data are plotted in blue, and fits to WAL theory (Eq.~\ref{HLN}) are plotted in yellow. (d) Coherence length \lphi\ and prefactor $\alpha$ extracted from Eq.~\ref{HLN} for each \vtg. Both \lphi\ and $\alpha$ are modulated by the applied gate voltage.}
\end{center}
\vspace{-0.8cm}
\end{figure}

From the WAL point of view, the surface and bulk states can be regarded as independent phase coherent channels as long as the carriers in one channel lose coherence before being scattered into the other. In this case, the majority of closed loops responsible for WAL will involve states from a single channel, and each channel will exhibit its own WAL correction. This condition can be formulated as $\tau_{SB} > \tau_\phi$, where $\tau_{SB}$ is the effective surface to bulk scattering time. In the opposite limit, $\tau_{SB} < \tau_\phi$, charge carriers scatter between the bulk and surface states while maintaining phase coherence, effectively becoming a single phase-coherent channel. We therefore interpret the gate-dependence of $\alpha$ as direct modulation of $\tau_{SB}$ via the formation of a depletion layer between the top surface and bulk carriers (see Fig.~2(c)), spatially separating them and suppressing the scattering probability between them. 

It is important to discuss the validity of Eq.~\ref{HLN} when the coupling between channels is tunable. In the fully decoupled regime, each channel \textit{i} has correction $\Delta G_{i}$ which follows Eq.~\ref{HLN} (with alpha = -0.5) and depends on $l_{\phi,i}$, yielding for the total correction $\Delta G_{tot} = \Delta G_1 + \Delta G_2$. Decomposing Eq.~\ref{HLN} into its logarithmic and digamma components~\cite{SI}, we note that the latter approaches a constant value for $B > B_o \approx 10-20$mT. Since our data extend to a few T, where the change in $\Delta G$ is dominated by the logarithmic component, Eq.~\ref{HLN} is a very good approximation to $\Delta G_{tot}$, with $\alpha = -1$ and an effective coherence length $l_{\phi}^{eff} = \sqrt{l_{\phi1}l_{\phi2}}$, as shown in~\cite{SI}. As a consistency check, we can find the dataset where Eq.~\ref{HLN} yields $\alpha = -1$, and fit the data to $\Delta G_{tot}$~\cite{SI}. We find the fit to agree very well with the data both for $B > B_o$ and $B < B_o$, with coherence lengths $l_{\phi 1,2}$ = 135, 77nm. When $-0.5 > \alpha > -1.0$ the physical interpretation of the parameters extracted from Eq.~\ref{HLN} is not trivial and no theoretical model exists to predict the \mc\ in the crossover coupling region. The evident success of the fit indicates that the logarithmic correction is robust, and $\alpha$ can be used phenomenologically as a measure for channel separation.

We have so far encountered two effective regimes differing in the degree of channel separation: (i) \vtg\ = -8V, where $\alpha \sim -1 = -(1/2+1/2)$, indicating a decoupling of the top surface from the rest of the system; (ii) \vtg\ = +6V, where $\alpha \sim -0.7$, indicating that the top surface is only partially decoupled. In highly doped samples  (Device C, Fig.~4(b)) we find a third regime, where $\alpha$ = -0.5, \RH\ = -3.7 $\Omega/T$, and \textit{G} = 57 \esh. This result is consistent with other WAL studies~\cite{MBE_WAL,Chen_GateWAL_2010} and suggests that the surface and bulk channels in highly doped TIs are fully coupled. This is consistent with our model, since for \vtg\ $\ge$ 0 the surface and bulk states co-exist in space and $1/\tau_{SB}$ should depend on the momentum difference between the surface and bulk bands, which become closer as the density increases~\cite{TI_ARPES,Park_ARPES}.


\begin{figure}
\label{Figure_4}
\begin{center}
\includegraphics[width = 78mm]{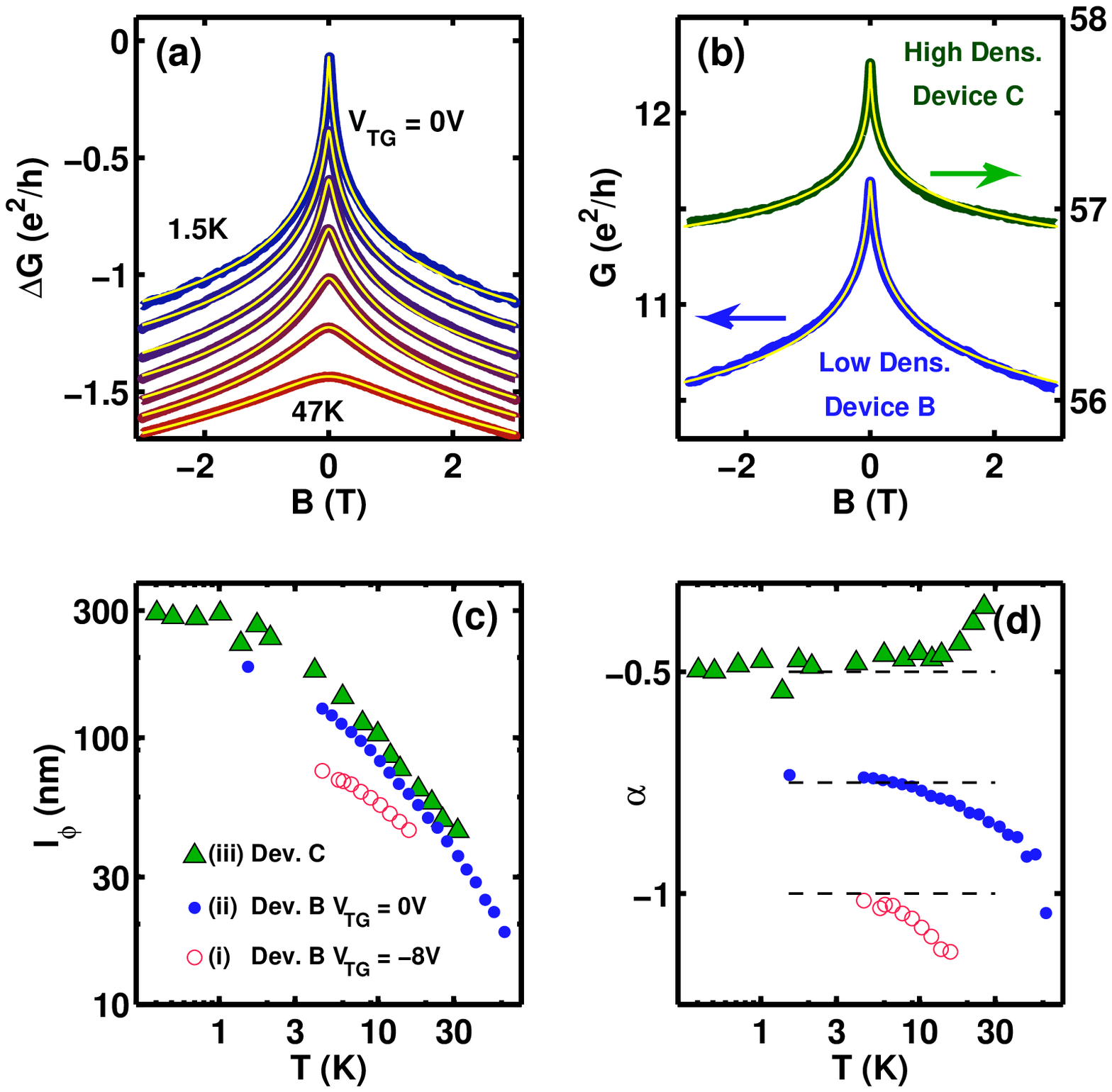}
\vspace{-0.8cm}
\caption{(a) Magnetoconductance \delG\ measured on low-density Device B at \vtg\ = 0V for different temperatures (from 1.5K to 47K). The curves are vertically shifted for clarity. Superimposed on each scan, in yellow, is the fit to Eq.~\ref{HLN}. (b) WAL traces and fits to Eq.~\ref{HLN} taken on low density Device B and high density Device C. Device B: \lphi\ = 175nm, $\alpha$ = -0.75; Device C: \lphi\ = 300nm, $\alpha$ = -0.5. (c) \lphi\ vs. $T$ for the three regimes discussed in the text. Open circle: Device B at the resistance peak (i); Full circle: Device B at \vtg\ = 0V (ii); Triangle: Device C (iii). (d) $\alpha$ vs. \textit{T}, for all three regimes.}
\end{center}
\vspace{-1.0cm}
\end{figure}

The effective number of phase coherent channels reflected in the magnitude of  $\alpha$ depends on the ratio $\tau_{SB}$ to \tauphi, which was controlled above by modulating $\tau_{SB}$ via the electric field effect. An independent control over this ratio can be obtained by varying the temperature, which directly controls \tauphi\ through dephasing. Figure~4 shows the change in the WAL correction with temperature in the different mixing regimes discussed above. Fig.~4(a) shows $\Delta G(B)$ in regime (ii) from $T$ = 1.5K to 47K, together with the corresponding fits to Eq.~\ref{HLN}. Similar data and fits are obtained for regimes (i) and (iii). The corresponding temperature dependence of $\alpha$ and \lphi\ is shown in Figs.~4(c) and (d). $\alpha$ decreases with temperature for regimes (ii) (from -0.75 to~$\sim$~-1) and (i) (-1 to -1.15), suggesting increased channel separation with temperature. This may seem surprising, since naively one expects larger channel mixing at higher temperatures. However, the behavior of $\alpha(T)$ is consistent with our model, and can be understood once the temperature dependence of \lphi\ is examined.  In Fig.~4(c) we see that \lphi\ is strongly temperature dependent, and since $\tau_\phi \sim  l_\phi^2$, this indicates that \tauphi\ decreases rapidly as $T$ is increased, which should result in a decrease in the $\tau_\phi/\tau_{SB}$ ratio, and consequently an increase in the channel separation. The validity of this model requires that $\tau_{SB}$ changes slower than \tauphi. This is expected in view of recent mobility data~\cite{Steinberg_2010} suggesting that the impurity scattering rate, which should govern $\tau_{SB}$, is indeed nearly temperature independent below 40K. The results in regime (i), where $\alpha$ becomes smaller than -1, suggest that, in addition to the top-bulk separation identified above, also the bottom surface could be decoupling from the bulk . At very high densities (Device C), this behavior is not observed, likely because $\tau_{SB}$ is much shorter than \tauphi\ in the temperature range explored. The corresponding increase in $\alpha$ at high $T$ could be consistent with coherent transport being dominated by the bulk, where WAL is suppressed when $\tau_\phi \approx \tau_{SO}$.

During the preparation of this manuscript we became aware of work by Chen et al., ~\cite{Chen_2011_WAL}, which report a similar modulation of $\alpha$, but associate it with changes in the coherence lengths of electron and hole channels. Such interpretation cannot explain our data, since our samples are thicker and the bottom surface density is not tunable, nor can it explain our observed temperature dependence nor the fact that $\alpha$ = -0.7 at positive gate voltages.
 
We are grateful to L. Fu, P. A. Lee, K. Michaeli, L. I. Glazman and A. Yacoby  for useful discussions. H.S.  acknowledges financial support from the Israeli Ministry of Science. J.-B.L. and J.S.M. thank support from NSF DMR grant 0504158 and ONR grant N00014-09-1-0177. P.J-H. acknowledges support from a Packard Fellowship. This work was performed in part at the NSF funded Harvard Center for Nanoscale Systems and MIT Center for Materials Science and Engineering.

\end{document}